\documentclass{article}
\usepackage{fleqn}
\usepackage{amssymb}
\usepackage{epsfig}
\usepackage{verbatim}

\begin{document}
\begin{center}
{\bf\Large  Establishment of thermodynamic equilibrium\\ in
cosmological model with an arbitrary acceleration}\\
Yu.G.Ignatyev\\
Kazan Federal University,\\ Kremlyovskaya str., 35,
Kazan 420008, Russia\\[12pt]

{\bf keywords}: Early Universe, Local Thermodynamic Equilibrium,
Relativistic Kinetics, Scaling, Cosmic Rays
\end{center}

\begin{abstract}
It has been built a strict mathematical model of the
thermodynamic equilibrium's establishment in the originally
nonequilibrium cosmological ultrarelativistic plasma for the
arbitrary accelerated Universe under the assumption that scaling of
interactions of elementary particles is restored at energies above
the unitary limit. It is shown, that in case of Universe positive
acceleration, thermodynamic equilibrium never can be completely
restored.
\end{abstract}

\section{Introduction}
In 1986 a hypothesis that particles interaction scaling is
restored at energies above the unitary limit \cite{Land_Unit} has
been put forward by the author; it was based on the results of the
analysis of axiomatic S-matrix theory and experimental data about
extra-high energy particles' cross-section of scattering. Author
also proposed a formula for the asymptotic cross-section of
scattering above the unitary limit which had scaling behavior and
was composed of 3 fundamental constants - \cite{Yu_1986}. A
restoration of the elementary particles scaling in range of an
extra-high energies is equivalent to the field equations'
conformal invariance restoration in the short-wave,
quasi-classical limit. The conformal invariance of the
relativistic kinetic theory in the ultra-relativistic limit is
also restored at that \cite{Yu_Conf}. Basic regulations of theory
of thermo\-dy\-na\-mic equilibrium's restoration in
ultrarelativistic cosmological plasma were formulated in
\cite{Yu_1986}; based on these regulations the numerical model of
this process at ultrarelativistic stage of the Universe expansion
was built and main results of the theory were formulated.
Developed in mentioned articles theory revealed a number of
interesting peculiarities of the process of thermodynamic
equilibrium resto\-ra\-ti\-on in cosmological plasma and allowed
to impose constrains on the parameters of the initial
distribu\-tion of particles. In particular, it was shown that
particles possessing energies more than $10^{12}\div 10^{13}$ Gev
in modern Universe can be relict ones carrying information about
cosmological singularity. In artic\-les \cite{Yu_1986},
\cite{Ign_Ziat1}
---\cite{Ign_Ziat3} it was formulated a diffused model and
investigated an evolution model of extra-high energy particles'
spectrum in ultrarelativistic Uni\-ver\-se. On the seminar
``Gracos'' (Yalchik, 2009) professor V.N.Melnikov put a question
to the author, which corrections to the Universe nonequilibrium
models can bring the factor of Universe acceleration that was
revealed in 1998 and became an acknow\-led\-ged fact nowadays. An
attempt to answer this question stretched for three years and
unexpectedly brought to the significant generalization and the
development of theory of Universe thermodynamic restoration as
well as detection of series of rather interesting facts. This
paper is devoted to that question's answer, for which author is
greatly thank\-ful to professor V.N.Melnikov.

\section{Common principles of energy-balance equation's construction and solution}
\subsection{Matter model}%
As is known, Einstein equations in case of isotropic homogeneous
cosmological model with zero three-dimensional curvature are reduced
to the system of two ordinary first-order differential equations:
\begin{equation}\label{IV.1}
\frac{\dot{a}^2}{a^2}=\frac{8\pi}{3}\varepsilon;
\end{equation}
\begin{equation}\label{IV.2}
\dot{\varepsilon}+3\frac{\dot{a}}{a}(\varepsilon+{\rm p}(\varepsilon))=0.
\end{equation}

Then:
\begin{equation}\label{IV.3}
\varepsilon=\varepsilon_p+\varepsilon_s;\quad {\rm p}={\rm p}_p+{\rm p}_s,
\end{equation}
where $\varepsilon_p,{\rm p}_p$  are energy density and pressure of the
cosmological plasma, $\varepsilon_s,{\rm p}_s$ are energy density and
pressure of various fundamental fields, probably, scalar ones,
leading to the Universe acceleration.

Invariant acceleration of the Universe
\begin{equation}\label{IV.4}
\Omega=\frac{a\ddot{a}}{\dot{a}^2}
\end{equation}
is associated with an  {\it effective barotropic factor} of matter,
$\varkappa\equiv {\rm p}/\varepsilon$, by relation:
\begin{equation}\label{IV.5}
\Omega=-\frac{1}{2}(1+3\varkappa).
\end{equation}
Thereby, during the acceleration stage ($t>t_1$):
\begin{equation}\label{IV.6}
0<\Omega<1, \Rightarrow -1 <\varkappa<-\frac{1}{3}.
\end{equation}
Until this moment  $\varkappa\in [1/3,-1/3]$. According to
(\ref{IV.1}) -- (\ref{IV.2}) scale factor and total energy density
 {\it at given constant barotropic factor} are changing by rule:
\begin{equation}\label{IV.7}
a=a_1 t^{2/3(\varkappa+1)};\; \varepsilon=\frac{1}{6\pi
(\varkappa+1)^2t^2}, \quad \varkappa+1\not=0
\end{equation}
Let us rewrite relations (\ref{IV.7}) in more convenient form using
relation (\ref{IV.5}):
\begin{equation}\label{IV.8}
a=a_1 t^{1/(1-\Omega)};\; \varepsilon=\frac{3}{8\pi
(1-\Omega)^2t^2}, \quad \Omega<1.
\end{equation}

Let us note that at any values of invariant acceleration $\Omega
={\rm Const}\in [-1,1)$ energy density is proportional to $t^{-2}$.
\subsection{General assumptions of the model}
Let us formulate {\it general assumptions}, which are laid in the
basis of the presented below mathematical model of thermodynamic
equilibrium's restoration in the expanding Universe:\\
\noindent\underline{$1^o$}. \hskip 12pt An asymptotic character of
elementary partic\-les interaction in range of extra-high
energies:
\par $1^o$a. A restoration of elementary particles' {\it scaling
of interactions} at energies above the {\it unitary limit}
\footnote{$s_u\simeq m_u^2$, where $m_u$ is a mass of the heaviest ,
``latest'', intermediate X-boson type particle participating in the
4-particle interactions}:
\begin{equation}\label{IV.10_0}
\left.\sigma\right|_{s>s_u}\sim \frac{1}{s},
\end{equation}
where $s$ is a first kinematic invariant -- the relativistic square of
a moment of two colliding particles, ``a'' and ``b'',  (see, e.g. \cite{Pilk}):
$$s=(p_a+p_b)^2\equiv g_{ik}(p^i_a+p^i_b)(p^k_a+p^k_b);$$
\par $1^o$b. {\it unification of all interparticle interactions}
on basis of fundamental constants $G,\hbar,c$ at energies above
the unitary limit, which subject to (\ref{IV.10_0}) leads to
formula of {\it universal asymptotic cross-section of scattering}
(details see in \cite{LTE1}):
\begin{eqnarray}\label{IV.11}
\sigma_0(s)=\frac{8\pi}{s\Lambda(s)},\\
\label{IV.12}%
\Lambda(s)=\ln^2\left(1+\frac{s_0}{s}\right),
\end{eqnarray}
$s_0=4$ is a squared total energy of two colliding Planck
masses\footnote{in the universal system of units, which is used in
this article $G=\hbar=c=1$, ; $s_0=4\hbar
c^5/G$} in the ordinary units.\\
\noindent\underline{$2^o$}. \hskip 12pt {\it Minimality of
fundamental fields' coupling with a cosmological plasma}. This
automatically means that ``energy conservation law'' (\ref{IV.2}) is
held separately for fundamental fields and plasma:
\begin{eqnarray}\label{IV.10}
\dot{\varepsilon_s}+3\frac{\dot{a}}{a}(\varepsilon_s+{\rm p}_s(\varepsilon_s))=0,\\%
\label{IV.11}
\dot{\varepsilon_p}+3\frac{\dot{a}}{a}(\varepsilon_p+{\rm p}_p(\varepsilon_p))=0.
\end{eqnarray}
\noindent\underline{$3^o$}. \hskip 12pt {\it Ultrarelativistic state
equation of cosmological plasma} at the stage of expansion that is
being researched:
\begin{equation}\label{IV.12}
{\rm p}_p=\frac{1}{3}\varepsilon_p,\quad (\ref{IV.11}) \Rightarrow
\varepsilon_pa^4={\rm Const}.
\end{equation}
\noindent\underline{$4^o$}. \hskip 12pt  {\it Ultrarelativistic
start of the Universe}:
\begin{equation}\label{IV.13}
\lim_{t\to 0}\varkappa=\frac{1}{3}.
\end{equation}

\vskip 12pt
In the works quoted above \cite {Yu_1986, Yu-Dim1, Yu-Dim2}\footnote {unlike
works \cite {Ign_Ziat1} - \cite {Ign_Ziat3}.} weak
viola\-tion of thermodynamical equilib\-rium was supposed in the Universe, when the greater part of particles, $n _ {ne} $, is in a thermal equilibrium state, and only for a minor part of particles, $n _ {ne} $, the thermal equilib\-rium is violated. As thermodynamic equilib\-rium at early stages of
the Universe can be broken only behind a unitary limit $E (p)> E_u $\footnote{$p $ is a
particle moment, $E (p) $ is its energy. As
opposed to the momentum variable, $p$, pressure is denoted by the
Roman type, -- ${\rm p}$.}, it leads to a following kind of function
of distribution of particles \cite {Yu_1986, Yu-Dim1}:
\begin{equation}\label{IV.13a}
\!\!f(p)\approx\!\! \left\{%
\!\!\begin{array}{ll}
f_0(p)=\!\!\left[\exp\left(\frac{-\mu+E_a(p)}{T} \right)\!\! \pm 1 \right]^{-1}\!\!\!, & p<p_u;\\[10pt]
\Delta f(p); f_0(p)\ll \Delta f(p)\ll 1, & p>p_u,
\end{array}\right.
\end{equation}
where $\mu(t)=\mu_a(t)$ are chemical potentials and $T(t)$ is the temperature of the equilibrium component of the plasma. Thus in the range $E>E_u \to s>s_u$ one can observe an anomalously great number of particles as compared with the equilibrium one, but still it will be small as compared with the total number of equilibrium particles. However, thus energy nonequilibrium components can be more energy equilibrium components.

It allows to consider plasma, as the system consisting of two subsystems: equilibrium with some temperature $T (t) $ and nonequilibrium, gradually losing energy at the expense of a transmission of energy equilibrium a component at collisions with equilibrium particles \cite {Yu_1986, Yu-Dim1}. Thus, in our model equilibrium a component it is warmed up by super\-thermal particles nonequilibrium components, that as a result leads to slower cooling of equilibrium plasma, than in standard model. This process is described {\it by the power balance equation} which the plasma component (\ref {IV.11}) is a direct consequence of the law of preservation of total energy) and will be formulated more low.

\subsection{Energy balance of the cosmological plasma}
From (\ref{IV.11}) subject to (\ref{IV.12}) right away follows:
\begin{equation}\label{IV.17}
\varepsilon_p a^4\equiv \tilde{\varepsilon}_p = {\rm Const},
\end{equation}
where $\tilde{\varepsilon}_p$ is a conformal energy density of the
cosmological plasma;
\begin{equation}\label{IV.17a}
\varepsilon_p = \varepsilon_e+\varepsilon_{ne},
\end{equation}
$\varepsilon_e$ is a energy density of the
equilibrium compo\-nents of plasma, and $\varepsilon_{ne}$ a energy density of the
non\-equilib\-rium components of plasma. As both compo\-nents of plasma,
equilibrium and non\-equilib\-rium, are ultra\-rela\-tivis\-tics, for them
conditions are separately satisfied:
 \begin{equation}\label{IV.17b}
 {\rm p}_e=\frac{1}{3}\varepsilon_e;\quad {\rm p}_{ne}=\frac{1}{3}\varepsilon_{ne}.
 \end{equation}

 Let us determine this constant, setting
according to (\ref{IV.8}) on the initial ultrarelativistic stage of
expan\-sion:
\begin{equation}\label{IV.18}
\left. a(t)\right|_{t\to 0}=\sqrt{t}.
\end{equation}
Then for the conformal energy density of plasma we obtain,
suggesting that cosmological plasma if the single ultrarelativistic
component of matter, ---
\begin{equation}\label{IV.19}
\tilde{\varepsilon}_p=\frac{3}{32\pi}.
\end{equation}
Let us then introduce temperature $T_0(t)$ of the cosmological
plasma in the ideal Universe, in which at this point in cosmological
time $t$ the whole plasma is locally balanced. Thus, plasma's energy
density is equal to
\begin{equation}\label{IV.20}
\varepsilon_p^0=\frac{N_0\pi^2}{15}T_0^4(t),
\end{equation}
where $N_0=N^0_B+1/2 N^0_F$ is an effective number of equilibrium
particles' types (bosons and fermions) in plasma with temperature
$T_0$.

Hence subject to (\ref{IV.19}) we obtain the evolution law of
plasma's temperature in the equilibrium Universe:
\begin{equation}\label{IV.21}
T_0(t)=\frac{1}{a(t)}\left(\frac{45}{32\pi^3 N_0}
\right)^{\frac{1}{4}}.
\end{equation}
With respect to $N_0$ --- {\it effective number of types} of
thermodynamically equilibrium particles, we suggest that $N_0(t)$ is
a slowly changing function of the cosmological time:
\begin{equation}\label{IV.22}
\dot{N}_0 t\ll 1.
\end{equation}

Let then $T(t)$ is a real temperature of equilibrium component of
cosmological plasma, and $\Delta f_a(p,t)$ is a distribution
function of ``a''-sort nonequilibrium plasma particles. Then we find
energy densities of the equilibrium, $\varepsilon_e $, and
nonequilibrium, $\varepsilon_{ne}$, components:
\begin{eqnarray}\label{IV.23}%
\varepsilon_e &=&\frac{N\pi^2}{15}T^4(t);\\
\label{IV.24}%
\varepsilon_{ne}&=&\frac{1}{2\pi^2}\sum\limits_a (2S+1)
\int\limits_0^\infty p^3\Delta f_a(p,t)dp,
\end{eqnarray}
where $N(t)=N_B+1/2 N_F$ is an effective number of equilibrium
particles' (bosons and fermions) types in plasma with temperature
$T(t)$. Expressing further with a use of (\ref{IV.21}) the scale
factor via temperature $T_0(t)$ and introducing new {\it
dimensionless} conformal momentum variable $\tilde{p}$:
\begin{equation}\label{IV.25}
p=\left(\frac{45}{32\pi^3}\right)^{\frac{1}{4}}\cdot\frac{\tilde{p}}{a(t)}=T_0(t)N_0^\frac{1}{4}\tilde{p},
\end{equation}
we obtain for (\ref{IV.24}):
\begin{equation}\label{IV.26}
\tilde{\varepsilon}_{ne}=\frac{45}{64\pi^5}\sum\limits_a (2S+1)
\int\limits_0^\infty \tilde{p}^3\Delta f_a(\tilde{p},t)d\tilde{p}.
\end{equation}
Next, from (\ref{IV.21}) and (\ref{IV.23}) we obtain for the
conformal energy density of the equilibrium component of plasma:
\begin{equation}\label{IV.26}
\tilde{\varepsilon}_e=\frac{3}{32\pi}y^4,
\end{equation}
where the dimensionless function is introduced, $y(t)$ is a -- {\it
relative temperature} \cite{Yu-Dim1}:
\begin{equation}\label{IV.27}
y(t)=\frac{T(t)}{T_0(t)}\leqslant 1.
\end{equation}
From (\ref{IV.26}) it is possible to obtain the following relation:
\begin{equation}\label{IV.28}
\sigma(t)\equiv y^4(t)\equiv
\frac{\varepsilon_e}{\varepsilon_p}\equiv
\frac{\tilde{\varepsilon}_e}{\tilde{\varepsilon}_e+\tilde{\varepsilon}_{ne}}.
\end{equation}

Thus, cosmological plasma's energy conservation law (\ref{IV.19})
with a use of (\ref{IV.24}) and (\ref{IV.26}) can be rewritten in
form:
\begin{equation}\label{IV.29}
y^4+\frac{15}{2\pi^4}\sum\limits_a (2S+1) \int\limits_0^\infty
\tilde{p}^3\Delta f_a(\tilde{p},t)d\tilde{p}=1.
\end{equation}
Relation (\ref{IV.29}) is called {\it plasma energy-balance
equa\-tion}. It is obtained with a usage of three model
suggestions
--- $2^o$, $3^o$, $5^o$. Let us note that in previous articles this
basic relation of mathematical model of ther\-mo\-dy\-na\-mi\-cal
equilibrium's restoration has been obtained under more special
suggestions. At given dependency of nonequilibrium particles'
distribution function on the temperature of plas\-ma's equilibrium
component and cosmological time the energy-balance equation
becomes a nonlinear integ\-ral equa\-tion relative to equilibrium
compo\-nent's tempe\-ra\-tu\-re. Therefore, to obtain this
equa\-tion in the explicit form it is necessary to solve the
kinetic equation for the nonequilibrium particles.

\section{Kinetic equation for nonequilibrium particles}
\subsection{Solution of kinetic equation}
As was shown in \cite{Yu-Dim1}, kinetic equation for
super\-ther\-mal particles in ultrarelativistic plasma has a form:
\begin{equation}\label{IV.30}
p\frac{\partial \Delta f_a}{\partial t}= -\frac{4\pi
N}{3}\frac{T^2(t)}{\Lambda(pT/2)}\Delta f_a,
\end{equation}
where $\Delta f_a=\Delta f_a(\tilde{p},t)$, $\Lambda=\Lambda(s)$.
Using here relation (\ref{IV.25}), we reduce (\ref{IV.30}) to form:
\begin{eqnarray}\label{IV.31}
\frac{\partial \Delta f_a}{\partial t}= & -{\displaystyle
\frac{8\pi N}{3\tilde{p}\Lambda(\frac{1}{2}\tilde{p}\ T_0T
N^{1/4})}}\times \nonumber\\  & {\displaystyle
\left(\frac{2\pi^3}{45}\right)^{1/4}}T^2(t)a(t)\Delta f_a.
\end{eqnarray}
Solving (\ref{IV.31}), obtain:
\begin{eqnarray}\label{IV.32}
\Delta f_a(t,\tilde{p})=\Delta f^0_a(\tilde{p})\times\nonumber\\
\exp\left[-\frac{8\pi}{3\tilde{p}}\left(\frac{2\pi^3}{45}\right)^{1/4}
\int\limits_0^t \frac{N aT^2 dt}{\Lambda(\frac{1}{2}\tilde{p}\
T_0T N_0^{1/4})}\right],
\end{eqnarray}
where
\begin{equation}\label{IV.33}
\Delta f^0_a(\tilde{p})\equiv \Delta f_a(0,\tilde{p}).
\end{equation}

\subsection{Transition to dimensionless normalized variables}
Let us introduce {\it an average conformal energy of
ultrarelativistic particles' nonequilibrium component at the initial
point in time},$\langle \tilde{p}\rangle_0$, --
\begin{equation}\label{IV.34}
\langle
\tilde{p}\rangle_0=\frac{\tilde{\varepsilon}(0)}{\tilde{n}(0)}\equiv
{\displaystyle%
\frac{\sum\limits_a (2S+1)\int\limits_0^\infty \Delta
f^0_a(\tilde{p})\tilde{p}^3d\tilde{p}}{\sum\limits_a
(2S+1)\int\limits_0^\infty \Delta
f^0_a(\tilde{p})\tilde{p}^2d\tilde{p}} %
}
\end{equation}
and {\it a dimensionless normalized momentum variable}, $\rho$, --
\begin{equation}\label{IV.35}
\rho\equiv \frac{\tilde{p}}{\langle\tilde{p}\rangle_0},
\end{equation}
so that
\begin{eqnarray}\label{IV.36}
\langle \rho\rangle_0=
\frac{\tilde{\varepsilon}(0)}{\langle
\tilde{p}\rangle_0\tilde{n}(0)}\equiv 1 \Rightarrow \nonumber\\
\langle \rho\rangle_0={\displaystyle%
\frac{\sum\limits_a (2S+1)\int\limits_0^\infty \Delta
f^0_a(\rho)\rho^3d\rho}{\sum\limits_a (2S+1)\int\limits_0^\infty
\Delta f^0_a(\rho)\rho^2d\rho}}=1.
\end{eqnarray}
According to nonequi\-lib\-rium plasma's mathematical model
particles' average energy in the initial nonequi\-lib\-rium
distribution must be greater and even much greater than particles'
thermal energy, thereby accor\-ding to (\ref{IV.25}),
(\ref{IV.34}) in the model being considered:
\begin{equation}\label{IV.37}
\langle \tilde{p}\rangle_0 \gg 1.
\end{equation}
Value $\langle \tilde{p}\rangle_0$ is in fact an {\it independent
parameter} of the model that is under consideration here and its
physical meaning is a relation of an average energy of
nonequi\-lib\-rium distribution particles to plasma temperature in
the equilibrium Universe in the initial point in
time\footnote{These values themselves are infinite but their ratio
is finite.}. As opposite to the conformal momentum variable
$\tilde{p}$ the average value of the dimensionless conformal
momentum variable $\rho$ in the initial distribution is
identically equal to $1$.

Let us transform an expression in the exponent (\ref{IV.32}), by
transition to dimensionless variables $y,\rho$. Taking into account
a weak dependency of the logarithmic factor $\Lambda$ on its
arguments and the decreasing character of the integrand in
(\ref{IV.32}), we receive the following evaluation of the
logarithmic factor:
\begin{equation}\label{IV.39}
\Lambda\left(\frac{1}{2}\tilde{p}\ T_0T N_0^{1/4}\right)\simeq
\Lambda(\langle\tilde{p}\rangle_0 T_0^2)\equiv \Lambda_0(t).
\end{equation}
Thus, with a logarithmic accuracy we represent a solution
(\ref{IV.32}) in compact form:
\begin{equation}\label{IV.40}
\Delta f_a(t,\rho)=\Delta f^0_a(\rho)
\exp\left(-\frac{2}{\rho}\int\limits_0^t \xi\frac{y^2}{a}dt
\right),
\end{equation}
where it is introduced the denotation:
\begin{eqnarray}\label{IV.41}
\xi\equiv\xi(t)=\left(\frac{5\pi}{18}\right)^{1/4}
\frac{N}{N_0^{1/2}\Lambda_0(t)}\nonumber\\
\approx 0.967
\frac{N}{N_0^{1/2}\Lambda_0(t)}
\approx\frac{N}{N_0^{1/2}\Lambda_0(t)}.
\end{eqnarray}
Introducing then a new {\it dimensionless time variable}, $\tau$,
--
\begin{equation}\label{IV.42}
\tau=2\int\limits_0^t \frac{\xi}{a}\ dt\ ,
\end{equation}
so that:
\begin{equation}\label{IV.43}
\frac{d\tau}{dt}\equiv 2 \frac{\xi}{a}>0,
\end{equation}
and a new {\it dimensionless function}, $Z(\tau)$, --
\begin{equation}\label{IV.44}
Z(\tau)=\int\limits_0^\tau y^2(\tau)d\tau ,
\end{equation}
we reduce the kinetic equation's (\ref{IV.40}) solution to form:
\begin{equation}\label{IV.45}
\Delta f_a(\tau,\rho)=\Delta f^0_a(\rho) \cdot{\rm
e}^{\displaystyle - \frac{Z(\tau)}{\rho}}.
\end{equation}
Let us investigate in details the {\it equation of coupling}
(\ref{IV.42}) between dimensionless variable $\tau$ and
cosmolo\-gi\-cal time $t$. Setting in (\ref{IV.42}) a power
dependence of the scale factor $a(t)$ on the cosmolo\-gi\-cal time
and taking into account a weak dependency of $\xi$ factor on time,
we obtain:
\begin{equation}\label{IV.46}
a\sim t^\alpha, \: (\alpha\not=1,0) \Rightarrow \tau\sim
t^{1-\alpha}; \: \alpha=1 \Rightarrow \tau\sim\ln t.
\end{equation}
Hence it follows that at $\alpha\leqslant 1 \to $
$\tau(\infty)=\infty$, and at $\alpha<1\to$
$\tau(\infty)=\tau_\infty<\infty$. Comparing relation (\ref{IV.46})
to relations (\ref{IV.5}) -- (\ref{IV.8}), we find:
\begin{eqnarray}\label{IV.47}
\varkappa\geqslant -\frac{1}{3} & (\Omega\leqslant0) & \Rightarrow
\tau(\infty)=+\infty;\\ %
\label{IV.48} %
\varkappa<-\frac{1}{3} & (\Omega>0) & \Rightarrow
\tau(\infty)=\tau_\infty<+\infty.%
\end{eqnarray}
Since the distribution function of cosmological plas\-ma's
nonequilibrium component (\ref{IV.45}) depends on time just by
means of {\it monotonously increasing function} $Z(\tau)$ of the
dimensionless time variable, relations (\ref{IV.47}) --
(\ref{IV.48}) mean that in ultrarelativistic cosmological plasma
in the Universe with a negative acceleration thermodynamical
equilibrium is reached asymptotically while {\it in the
accelerating Universe thermodynamical equilibrium is never
strictly reached}.

\subsection{Conformal energy density of the nonequilibrium component}
Substituting the solution of the kinetic equation in form
(\ref{IV.45}) in the expression for nonequilibrium particles'
conformal energy density, we obtain
\begin{equation}\label{IV.49}
\tilde{\varepsilon}_{ne}=\frac{45}{64\pi^5}\sum\limits_a (2S+1)
\int\limits_0^\infty \tilde{p}^3\Delta f^0_a(\rho) {\rm e}^{ -
\frac{Z(\tau)}{\rho}}.
\end{equation}
Let us perform an identity substitution on the given expression,
accounting that subject to the definition (\ref{IV.28})and
energy-balance equation (\ref{IV.29}):
\begin{equation}\label{IV.50}
\tilde\varepsilon_{ne}^0=(1-\sigma_0)\frac{3}{32\pi}:
\end{equation}
\begin{equation}\label{IV.51}
\tilde\varepsilon_{ne}\equiv
\frac{\tilde\varepsilon_{ne}}{\tilde\varepsilon_{ne}^0}\tilde\varepsilon_{ne}^0=(1-\sigma_0)\Phi(Z)\frac{3}{32\pi},
\end{equation}
where we introduced new {\it dimensionless function} $\Phi(Z)$
with a use of transformation to the dimen\-sion\-less momentum
variable $\rho$ (\ref{IV.35}) :
\begin{equation}\label{IV.52}
\Phi(Z)\equiv {\displaystyle \frac{\sum\limits_a (2S+1)
\int\limits_0^\infty \rho^3\Delta f^0_a(\rho) {\rm
e}^{\displaystyle - \frac{Z(\tau)}{\rho}}}%
{\sum\limits_a (2S+1) \int\limits_0^\infty \rho^3\Delta
f^0_a(\rho) }}.
\end{equation}

\subsection{Energy-balance equation solution and analysis}

As a result of definition (\ref{IV.44}) function $Z(\tau)$ satisfies
following conditions:
\begin{eqnarray}\label{IV.53}
Z'(\tau)=y^2(\tau)\Rightarrow Z'\ ^2=\sigma(\tau);\\
\label{IV.54}%
Z(0)=0;\quad Z'(0)=y^2(0)=\sqrt{\sigma_0},
\end{eqnarray}
where
\begin{equation}\label{IV.55}
Z'\equiv \frac{dZ}{d\tau}>0.
\end{equation}

Thus, subject to (\ref{IV.51}) -- (\ref{IV.53}) energy-balance
equation (\ref{IV.29}) can be rewritten in form of the
differen\-tial equation relative to function $Z(\tau)$:
\begin{eqnarray}\label{IV.56}
y^2+(1-\sigma_0)\Phi(Z)& =1\Rightarrow \nonumber\\
Z'^2+(1-\sigma_0)\Phi(Z)& =1,
\end{eqnarray}
solving which subject to relations (\ref{IV.54}) -- (\ref{IV.55}),
we find a formal solution in the implicit form:
\begin{equation}\label{IV.57}
\int\limits_0^Z \frac{du}{\sqrt{1-(1-\sigma_0)\Phi(u)}}=\tau.
\end{equation}

According to definition (\ref{IV.52}) function $\Phi(Z)$ is a
nonnegative one:
\begin{equation}\label{IV.58}
\Phi(Z)>0, \quad (Z\in [0,+\infty)),
\end{equation}
and
\begin{equation}\label{IV.59}
\Phi(0)=1;\quad \lim_{Z\to+\infty}\Phi(Z)=0.
\end{equation}
Calculating the first and the second derivatives of function
$\Phi(Z)$, by $Z$ and differentiating the relation (\ref{IV.52}) by
$Z$, obtain:
\begin{eqnarray}\label{IV.60}
\Phi(Z)'_Z<0, \quad (Z\in [0,+\infty));\\
\label{IV.61}%
\Phi(Z)''>0,\quad (Z\in [0,+\infty)).
\end{eqnarray}

In consequence of (\ref{IV.60}) function $\Phi(Z)$ is strictly
monotonously decreasing one but then as a result of the relations
(\ref{IV.59}) this function is limited on the interval:
\begin{equation}\label{IV.62}
\Phi(Z)\in [0,1];\quad (Z\in [0,+\infty)),
\end{equation}
and function $\Phi(Z)$ has a concave graph. As a result of these
properties of function $\Phi(Z)$ equation $\Phi(Z)=\Phi_0$ within
the limit being researched, always has a single and just only one
solution $Z=Z_0$, i.e. {\it mapping $Y=\Phi(Z)$ on the set of
nonnegative numbers is bijective}.

Next, from (\ref{IV.53}) it follows that function $Z(\tau)$
monotonously increases over the interval $\tau\in[0,\tau_\infty]$.
Differentiating relation (\ref{IV.56}) by $\tau$ like a composite
function, we find:
\begin{equation}\label{IV.63}
Z'[2Z''+(1-\sigma_0)\Phi'_Z]=0.
\end{equation}
Hence as a result of  $Z'$ positivity (\ref{IV.56}) let us find the
second derivative:
\begin{equation}\label{IV.64}
Z''=-\frac{1}{2}(1-\sigma_0)\Phi'_Z.
\end{equation}
Therefore in consequence of  (\ref{IV.60}) and(\ref{IV.27}) --
(\ref{IV.28}) we obtain from (\ref{IV.64}):
\begin{equation}\label{IV.65}
Z''>0,
\end{equation}
i.e. graph of $Z(\tau)$ function is also concave. Then,
differentiating (IV.53), subject to (\ref{IV.65}) find:
\begin{equation}\label{IV.66}
y'>0,
\end{equation}
--- i.e. function $y(\tau)$ (and function $\sigma(\tau)$ together with it)
is a monotonously increasing one. From the other hand it is limited
from below by the initial value $y_0$ ($\sigma_0$), and from above
by value $1$:
\begin{equation}\label{IV.66}
y'>0,y\in [y_0,1);\quad \sigma'>0, \sigma\in [\sigma_0,1).
\end{equation}

Listed properties of functions $y(\tau)$, $Z(\tau)$ and $\Phi(Z)$
assert the bijectivity of chain of mappings $\tau \leftrightarrow
y$, $y \leftrightarrow Z$, $Z \leftrightarrow \Phi$. Finally, each
value $\Phi$ has a one and only one corresponding value $Z$ and one
and only one value $\tau$: $\tau\leftrightarrow \Phi$. To close this
chain it is enough to determine functions $y(\tau)$ and $Z(\tau)$
coupling by means of the energy-balance equation(\ref{IV.57}):
\begin{equation}\label{IV.67}
y=[1-(1-\sigma_0)\Phi(Z)]^{1/4}.
\end{equation}

Equations (\ref{IV.57}) and (\ref{IV.67}) are the {\it parametric
solution of the energy-balance equation} (\ref{IV.56}), and above
mentioned properties of functions $\Phi(Z)$ and $Z(\tau)$ assert
the {\it uniqueness of the solution}. Accor\-ding to (\ref{IV.52})
function $\Phi(Z)$ is fully determined by the initial distribution
of nonequilibrium particles $\Delta f^0_a(\rho)$. Therefore from
the mathematical point of view the problem of thermodynamical
equilibrium recreation in Universe with arbitrary acceleration is
completely solved. Concrete models are determined by dark matter
model and model of initial nonequi\-lib\-rium distribution of
particles.

Let us differentiate now the relation (\ref{IV.64}) by $\tau$ and
take account of the association (\ref{IV.53}) between  $y(\tau)$ and
$Z(\tau)$:
\begin{eqnarray}
Z'''=-\frac{1}{2}(1-\sigma_0)\Phi''_{ZZ}Z'\nonumber\\
\Rightarrow
y''y=-y'^2-\frac{1}{4}(1-\sigma_0)\Phi''_{ZZ}y^2.\nonumber
\end{eqnarray}
Thus, as a result of (\ref{IV.61}):
\begin{equation}\label{IV.68}
y''<0,
\end{equation}
--- i.e. graph of function $y(\tau)$, and
$\sigma(\tau)$ graph together with it, are concave. Then since
$\Phi_Z(Z\to\infty)=0$, from (\ref{IV.64}) it follows:
\begin{equation}\label{IV.69}
\lim_{\tau\to\infty}y'(\tau)=0 \Rightarrow
\lim_{\tau\to\infty}\sigma'(\tau)=0,
\end{equation}
--- i.e., value $\sigma=1$ is reached asymptotically at
$\tau\to\infty$. This allows to draw a qualitative graph of
functions $y(\tau)$ (Fig. \ref{ris0}). Finiteness of the
dimensionless time $\tau_\infty$ conducts to the establishment of
the limiting value of function $y(\tau)$:
\begin{equation}\label{IV.70}
y(\tau_\infty)=y_\infty <1 \Rightarrow
\lim_{t\to\infty}y(t)=y_\infty<1.
\end{equation}
In consequence of that the certain part of comsolo\-gi\-cal plasma
energy is forever conserved in the nonequi\-lib\-rium superthermal
component:
\begin{equation}\label{IV.71}
\lim_{t\to\infty}\frac{\varepsilon_{ne}(t)}{\varepsilon_p(t)}=1-\sigma_\infty
= \left\{\begin{array}{ll}
=0, & \tau_\infty=\infty\\
>0, &  \tau_\infty<\infty
\end{array}\right.
.
\end{equation}
According to (\ref{IV.47}) -- (\ref{IV.48}) it is possible just for
the accelerated expanding Universe.

\section{Exact model of transition from the ultrarelativistic stage to the inflationary one}
Let us consider a simple model of matter consisting of 2
components -- minimally coupled massive scalar field (cosmological
member) with a state equation:
\begin{equation}\label{V.13}
p_s=-\varepsilon_s,
\end{equation}
and a ultrarelativistic plasma with a state equation
(\ref{IV.12}). Then an overall barotropic factor and invariant
acceleration can be written in form:
\begin{equation}\label{V.13}
\varkappa(t)=\frac{1}{3}\frac{1-3\delta}{1+\delta};\quad
\Omega(t)=-\frac{1-\delta}{1+\delta},
\end{equation}
where
\begin{equation}\label{V.13}
\delta=\delta(t)=\frac{\varepsilon_s}{\varepsilon_p} .
\end{equation}
\begin{center}
\parbox{8cm}{%
\refstepcounter{figure}\includegraphics[width=8cm,
height=5cm]{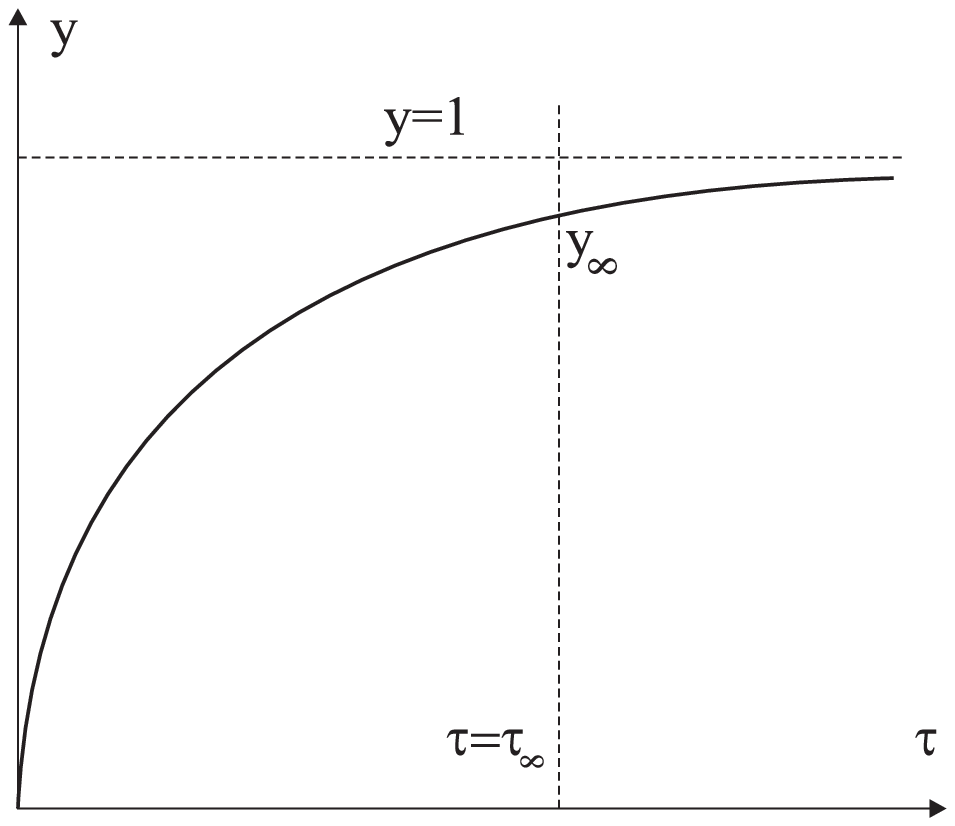} \noindent {\bf\small Fig. \thefigure.}\quad
{\small \label{ris0}Qualitative view of function $y(\tau)$
graph.}}\end{center}

Thus at $\delta={\rm Const}$ formulas (\ref{IV.7}) can be written in
the following convenient form:
\begin{equation}\label{V.14}
a=a_1 t^{(1+\delta)/2};\; \varepsilon=\frac{3}{32\pi}
\frac{(1+\delta)^2}{t^2}, \quad \Omega<1.
\end{equation}

Energy conservation laws(\ref{IV.10}) -- (\ref{IV.11}) take form:
\begin{eqnarray}\label{V.30}
\varepsilon_s={\rm Const}=\frac{3\Xi^2}{8\pi};\\%
\label{V.18}%
\varepsilon_p a^4 \equiv \tilde{\varepsilon}_p={\rm Const}\simeq
\frac{3}{32\pi}.
\end{eqnarray}
Substituting (\ref{V.30})-(\ref{V.18}) into equation (\ref{IV.1})
and integra\-ting it, we obtain:
\begin{eqnarray}\label{V.19}
a(t)=\frac{1}{\sqrt{2}}\left[
\left(t_0+\sqrt{t^2_0+b^2}\right){\rm e}^{(t-t_0)/2\Xi}-\right.\nonumber \\
\left. \frac{b^2}{t_0+\sqrt{t^2_0+b^2}}{\rm e}^{-(t-t_0)/2\Xi}
\right]^{\frac{1}{2}},
\end{eqnarray}
where:
\begin{equation}\label{V.20}
b^2=\frac{3}{32\pi\Xi^2}.
\end{equation}
Hence we have, in particular, for the scale factor at $t_0=0$ :
\begin{equation}\label{V.21}
a(t)=\frac{1}{\Xi}\sqrt{\frac{3}{32\pi}{\rm sh} \frac{t}{2\Xi}}
\end{equation}
Calculating according to (\ref{V.13}), (\ref{V.30}), (\ref{V.18})
and (\ref{V.21}) relation $\delta$, we find:
\begin{equation}\label{V.24}
\delta(t)=\left(\frac{3}{16\pi\Xi}\ {\rm sh}\frac{
t}{2\Xi}\right)^2.
\end{equation}
Next according to (\ref{V.13}) it is possible to calculate an
effective barotropic factor and an invariant accele\-ra\-tion (see
Fig. \ref{ris1}).

\begin{center}
\parbox{8cm}{%
\refstepcounter{figure}
\includegraphics[width=5cm,height=8cm,angle=-90]{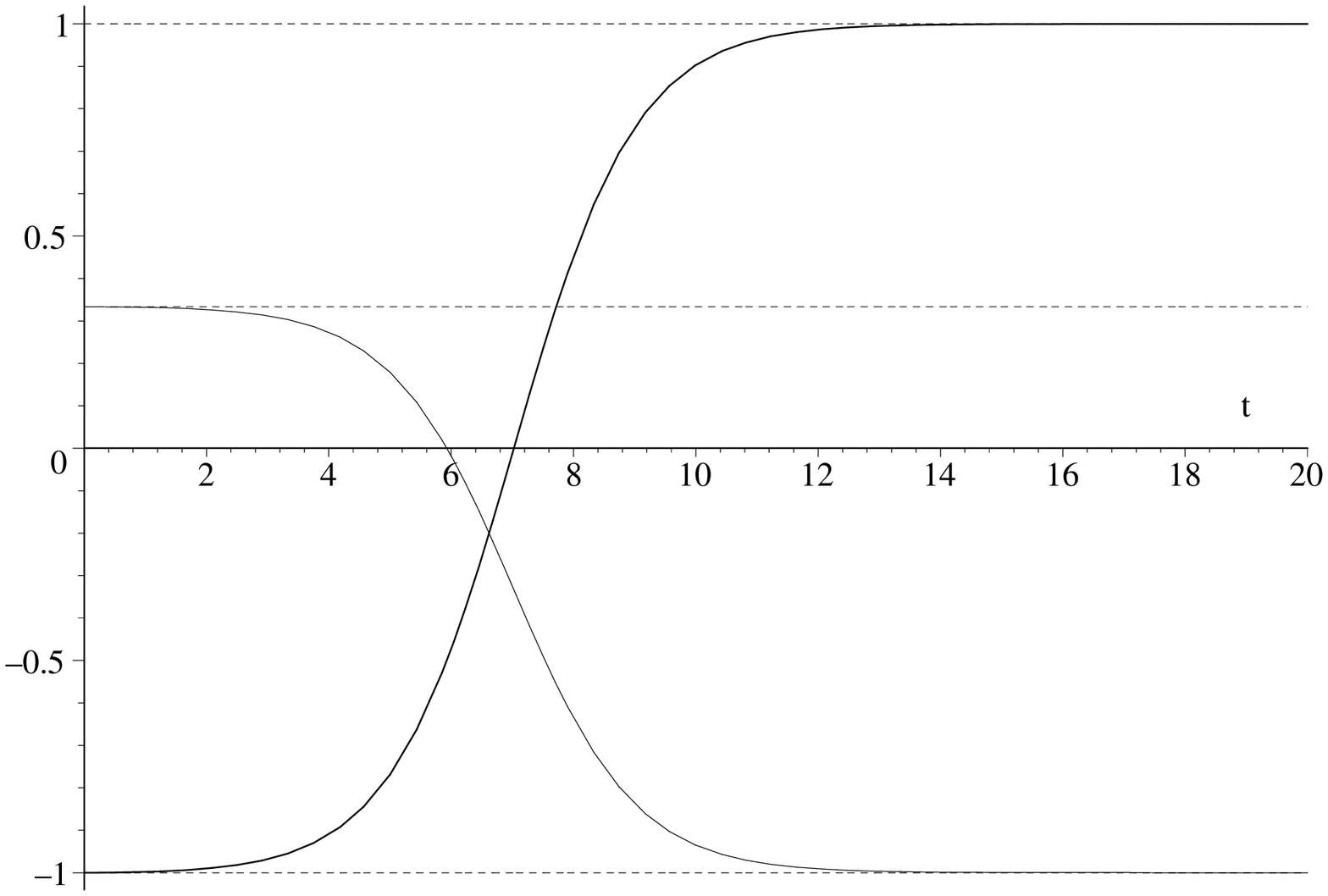}
\vskip 2pt \noindent {\bf\small Fig. \thefigure.}\quad {\small
\label{ris1}Evolution of the effective barotropic factor
$\kappa(t)$ (thin line) and invariant acceleration $\Omega(t)$
(heavy line) relative to the exact solution (\ref{V.21}) at
$\Xi=1$. Asymptotes $-1; \; 1/3; \;1$ are denoted by the dotted
lines.}}
\end{center}

Following figure shows that by means of para\-me\-ter $\Xi$ it is
easy to control the time of transition to the inflationary
acceleration regime $\kappa\to -1$. Let us recall that
cosmological time $t$ is measured in Planck units.

Thus according to (\ref{IV.42}) let us determine new dimensionless
time variable, $\tau$:
\begin{equation}\label{V.31}
\tau=\frac{2\Xi\langle\xi\rangle}{\langle\tilde{p}\rangle_0}{\rm F}(\varphi,1/\sqrt{2}),
\end{equation}
where:
\begin{equation}\label{V.32}
\varphi=\arccos \frac{1-{\rm sh}\; t/2\Xi}{1+{\rm sh}\; t/2\Xi};
\end{equation}
${\rm F}(\varphi,k)$ is an elliptic integral of the first type (see
e.g. \cite{Grad}):
\begin{equation}\label{V.34}
{\rm F}(\varphi,k)=\int\limits_0^\varphi
\frac{d\alpha}{\sqrt{1-k^2\sin^2\alpha}};\quad (k^2<1).
\end{equation}

Thus:
\begin{equation}\label{V.35}
\frac{d\tau}{dt}=\frac{1}{2\Xi}\frac{1}{{\rm
sh}\;\varphi}\frac{{\rm ch}\;\varphi }{1+{\rm sh}\;\varphi}>0;
\quad  \tau\in [0,\tau_\infty),
\end{equation}
where\footnote{$F(1,1/\sqrt{2}\approx 1.083216773$.}
\begin{equation}\label{V.36}
\tau_\infty=\lim\limits_{t\to+\infty}\tau(t)=\frac{2\Xi\langle\xi\rangle}{\langle\tilde{p}\rangle_0}
{\rm F}(1,1/\sqrt{2}).
\end{equation}

\begin{center}
\parbox{8cm}{%
\refstepcounter{figure}
\includegraphics[width=5cm,height=8cm,angle=-90]{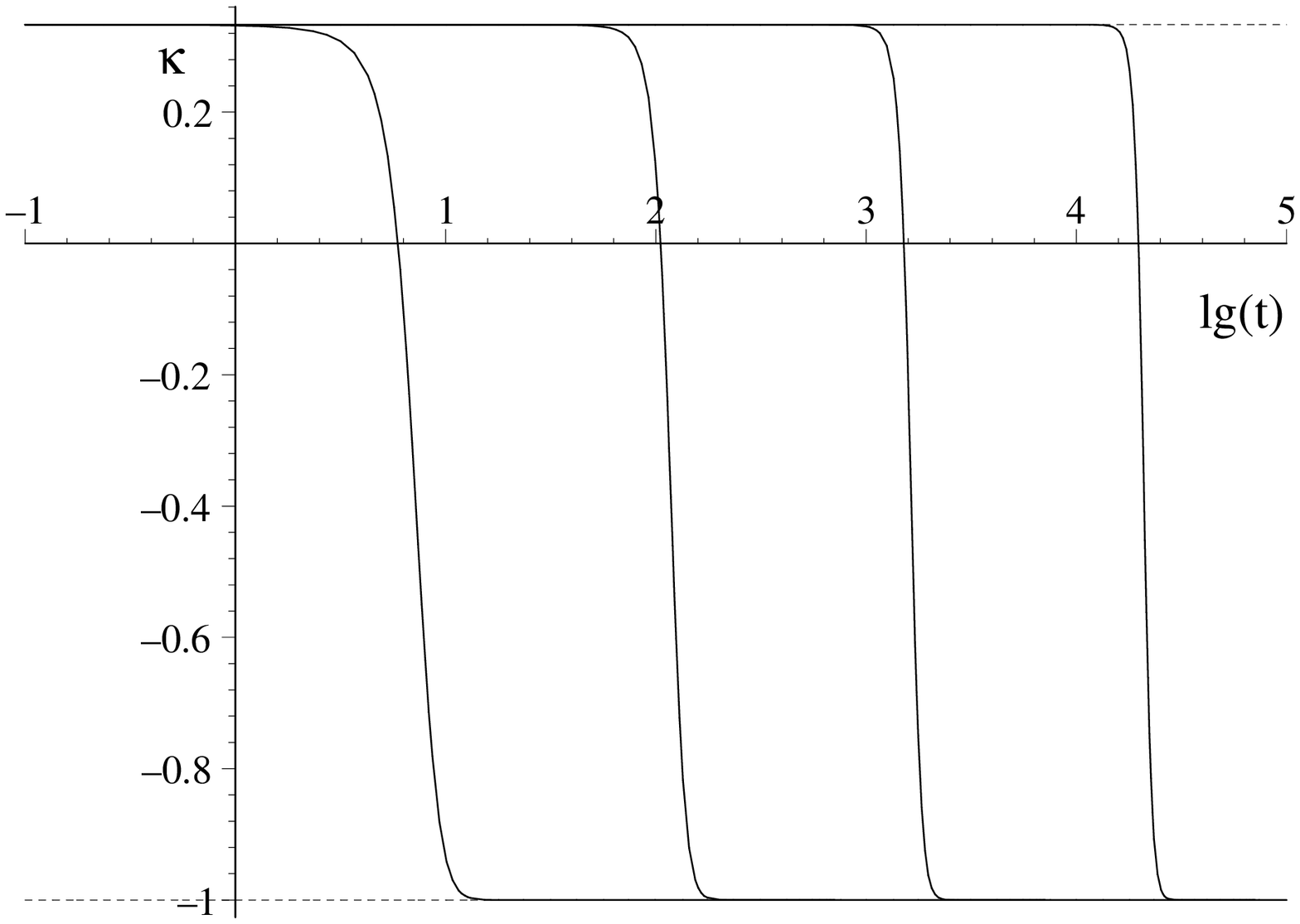}
\vskip 2pt \noindent {\bf\small Fig. \thefigure.}\quad {\small
\label{ris2}Evolution of the effective barotropic factor
$\kappa(t)$ at $\Xi=1;\;10;\;100;\;1000$ (left to right) relative
to the exact solution (\ref{V.21}). Dotted lines denote asymptotes
$\varkappa=-1;\;\varkappa=1/3$. Along the abscissa axis values
${\rm log}_{10}\; t$ are put.}}
\end{center}

\renewcommand{\refname}{\rm References}


\begin{thebibliography}{30}
\bibitem{Yu_1986}
Yu.G. Ignat'ev, J. Sov. Phys. (Izv. Vuzov). {\bf 29}, No 2, 19 (1986).
%
\bibitem{Yu_Conf}
Yu.G. Ignat'ev, J. Sov. Phys. (Izv. Vuzov). {\bf 25}, No 4, 75 (1982).
%
\bibitem{Land_Unit}
L.D. Landau, J. Sov. Phys. (JETP) {\bf 10}, 718 (1949).
%
\bibitem{LTE1}
Yu.G. Ignatyev, Grav. Cosmol., {\bf 13}, 31 (2007).
%
\bibitem{Yu-Dim1}
Yu.G. Ignatyev and D.Yu. Ignatyev, Grav. Cosmol., {\bf 13}, 31 (2007).
%
\bibitem{Yu-Dim2}
Yu.G. Ignatyev and D.Yu. Ignatyev, Grav. Cosmol., {\bf 14}, 309 (2008).
%
\bibitem{Ign_Ziat1}
Yu.G. Ignatyev and  R.A. Ziatdinov, Grav. Cosmol.,
{\bf 12}, No 4, 293 (2006).
%
\bibitem{Ign_Ziat2}
Yu.G. Ignatyev and  R.A. Ziatdinov, Grav. Cosmol.,
{\bf 14}, No 4, 301 (2008).
%
\bibitem{Ign_Ziat3}
Yu.G. Ignat'ev and  R.A. Ziatdinov, Russian Physics J., {\bf 52}, No 2, 210 (2009)
%
\bibitem{Pilk}
 H.M. Pilkuhn, Relativistic Particle Physics. 1979. New-York:
Springer-Verlag. 544 p.
%
\bibitem{Grad}
I.S. Grandshtein, I.M. Rizhik, Tables of Integrals, Sums and
Series. 1963. Moskow. Fizmatgiz.  1100 p. (In Russian).
\end{thebibliography}
\end{document}